\documentclass[twocolumn]{jpsj3}
\usepackage{txfonts}
\usepackage{amsmath, amssymb, braket, theorem, cite, stmaryrd,color}

\newcommand{\pr}[1]{#1^{\prime}}
\newcommand{\dg}[1]{#1^{\dag}}
\newcommand{\del}{\partial}
\newcommand{\vc}[1]{\boldsymbol #1}
\newcommand{\sgn}{\mathrm{sgn}}
\newcommand{\supp}{\mathrm{supp}}
\newcommand{\mfrak}[1]{\mathfrak{#1}}
\newcommand{\mcal}[1]{\mathcal{#1}}
\newcommand{\mbb}[1]{\mathbb{#1}}
\newcommand{\mrm}[1]{\mathrm{#1}}

\theoremstyle{plain}
\theorembodyfont{\upshape}
\newtheorem{thm}{Theorem}[section]
\newtheorem{prop}[thm]{Proposition}

\title{Perturbative Approach to Superfluidity under Nonuniform Potential}

\author{Shinji Koshida$^{1}$\thanks{koshida@vortex.c.u-tokyo.ac.jp}, Yusuke Kato$^{1}$}
\inst{$^{1}$Department of Basic Science, The University of Tokyo, \\3-8-1, Komaba, Meguro, Tokyo 153-8902, Japan} 

\abst{A perturbative way to investigate superfluid properties of various systems under nonuniform potential is presented.
We derive the perturbation expansion of the superfluid fraction, which indicates how liquid exhibits nonclassical rotational inertia, 
in terms of the strength of nonuniform potential
and find that the coefficient of the leading term reflects the density fluctuation of the system.
Our formulation does not assume anything about Bose-Einstein condensation
and thus is applicable to wide variety of systems.
Superfluid properties of some examples including (non-)interacting Bose systems, especially Bose gas in the mean field limit, 
(non-)interacting Fermi sytems, Tomonaga-Luttinger liquid and spinless chiral $p$-wave superfluid
are investigated.}

\begin{document}
\maketitle


\section{Introduction}
\label{sec:introduction}
What a system exhibits superfluidity is a longstanding problem in the condensed matter physics
from the first observation of frictionless flow in liquid $^{4}$He in 1938~\cite{Kapitza1938, AM1938}.
Much effort has been made to understand this phenomenon and we have seen great advance in this field of study~\cite{Leggett1973,Leggett1999,Leggett2006,LSSY2005},
but the complete criterion to superfluidity is still absent because of its complexity.
Then we try to give an approach to this problem in the present paper.
Before we start our discussion, we need to clarify problems concerning superfluidity and zoom in to the aspect we focus on
since superfluidity is a complicated and composite phenomenon.

First of all we have to distinguish between superfluidity of the ground state (or equilibrium state) and that of meta-stable states.
The latter is represented by existence of persistent current, macroscopic mass flow that is not accompanied by dissipation.
A stability condition for persistent current is given by Landau's criterion,
which says that a system can have relative motion to its container with velocity up to the critical one determined by the excitation spectrum,
or equivalently, a system with nonzero critical velocity can support persistent current.
Thus the critical velocity gives a conceptual criterion to superfluidity of meta-stable states,
and mechanism of (breakdown of) superfluidity has been studied by several authors~\cite{FPR1992, Hakim1997, KK2015}
being associated with this concept, but we do not expose them any more.
What we focus on in the present paper is the former one, superfluidity of the ground state.

Superfluidity of the ground state is characterized by nonclassical rotational inertia (NCRI).
Imagine water in a slowly rotating bucket with angular velocity $\omega$.
In the nonequilibrium stationary state, we can fairly expect that the water rotates with the same angular velocity
and its angular momentum has the form $I_{\mrm{cl}}\omega+O(\omega^{2})$,
where $I_{\mrm{cl}}$ is the moment of inertia computed from the mass density distribution of the content
assuming it were a rigid body.
If in the bucket is superfluid helium, we see things going differently.
The angular momentum again has the form $I\omega+O(\omega^{2})$,
but now the coefficient $I$ of $\omega$ is smaller than $I_{\mrm{cl}}$,
implying that there is a component staying stationary in the rotating bucket.
This property observed in superfluid is called NCRI and is widely accepted as characterization of superfluidity of the ground state.
NCRI is also observed experimentally as Hess-Fairbank effect~\cite{HF1967}.
Difference of superfluidity of the ground state from that of meta-stable states is in that
we always focus on the asymptotic behavior at $\omega\to 0$ to discuss superfluidity of the ground state.

We then replace our original problem by a problem: what a system exhibits NCRI?
Although there are some examples that are proved to exhibit NCRI~\cite{LSY2002,  KMSY2015} in the ground state,
we shall seek a general answer to this problem.
One answer is given by the concept of Bose-Einstein condensation (BEC).
The story goes as follows.
We characterize BEC through the formulation first introduced by Penrose and Onsager~\cite{PO1956}.
It says a system exhibits BEC if its one-particle density matrix has a maximum eigenvalue of macroscopic order.
The corresponding eigenfunction is called a condensate wave function, which describes macroscopic behavior of the system
and especially gives the superfluid velocity field.
From the fact that the wave function is single-valued,
the vorticity of condensate has to be quantized,
which leads to that the condensate in a slowly rotating container does not rotate.
This property is nothing but NCRI.
Although BEC seems to imply NCRI in the above discussion,
the quantization of vorticity is valid under assumption that the condensate wave function does not vanish
on any closed contour.
This assumption actually breaks down when we consider a system under presence of strong random potential,
and thus in such a situation there realizes BEC without superfluidity in the ground state~\cite{KMSY2015}.

It is also known that BEC is not necessary to superfluidity.
Actually, the low temperature phase of Kosterlitz-Thouless (KT) transiton~\cite{KT1973} exhibits superfluidity without BEC.
Thus BEC does not necessarily describe all physics of superfluidity
and we are led to investigate superfluidity of the ground state without assuming BEC.

We begin our study from a trivial observation that
any system seemingly exhibits NCRI if there is no obstruction or
the container has complete rotational symmetry.
Of course this does not mean that any system is regarded as superfluid
but just reflects the fact that a rotating container is the same as stationary one.
Thus we have to discuss superfluidity in terms of response to nonuniformity.
There are some theoretical approaches to superfluidity that is applicable to a system with nonuniform potential
such as discussions by Leggett~\cite{Leggett1970,Leggett1998},
but they require us to know the ground state wave function under nonuniform potential, which is not so easy.
Although superfluid properties of Bose-Einstein condensate under random potential are also investigated\cite{ABCG2002, KT2002},
as explained above, we have to study superfluidity without assuming BEC.
Thus our task is to determine what a system without nonuniform potential can support NCRI
when nonuniform potential is added to the system in a fully quantum mechanical way.

In the present paper, we construct the perturbation expansion of
the superfluid fraction, which indicates the extent to which the system exhibits NCRI
in terms of the strength of nonuniform potential.
Then we will find the leading contribution to the superfluid fraction is determined by density fluctuation of the system.
We do not assume BEC in the ground state, thus our perturbation theory
is applicable to wide variety of systems.

This paper is organized as follows.
In Sect. \ref{sec:definition}, we define the setting of our problem
and introduce the superfluid fraction.
In Sect. \ref{sec:perturbation}, we construct the perturbation expansion of the superfluid fraction
in terms of the strength of nonuniform potential.
We also rewrite the coefficient of the leading term using the dynamic structure factor,
which characterizes the density fluctuation of the system.
In Sect. \ref{sec:bounds}, we show an upper and lower bounds for the superfluid fraction
at perturbative level estimating the perturbation coefficient.
In Sect. \ref{sec:examples}, based on the results obtained in Sects. \ref{sec:perturbation} and \ref{sec:bounds},
we investigate superfluid properties of some examples such as
(non-)interacting bosons, especially Bose gas in the mean field limit, 
(non-)interacting fermions, Tomonaga-Luttinger liquid (TLL) and spinless chiral $p$-wave superfluid.
Finally in Sect. \ref{sec:conclusion}, we make conclusion and discussion.
Some computational details are described in Appendices.

Throughout this paper, we set $\hbar=1$.

\section{Definition of problem}
\label{sec:definition}
In this section, we clarify the setting of our problem and define the superfluid fraction of a system,
on which we focus to discuss superfluid property of the system.

To make our discussion general in this section,
we consider a system of $N$ particles in $d$-dimension described by a unital $\mbb{C}$-algebra
generated by symbols $x_{j}^{\mu}$ and $p_{j}^{\mu}$ ($j=1,2,\cdots ,N$, $\mu=1,2,\cdots, d$) with relations
\begin{equation}
	\label{eq:CCR}
	\left[x_{j}^{\mu},x_{k}^{\nu}\right]=\left[p_{j}^{\mu},p_{k}^{\nu}\right]=0,\ 
	\left[x_{j}^{\mu},p_{k}^{\nu}\right]=i\delta_{jk}\delta_{\mu\nu}
\end{equation}
for all $j,k=1,2,\cdots, N$ and all $\mu,\nu=1,2,\cdots, d$.
Here $\delta_{jk}$ and $\delta_{\mu\nu}$ denote the Kronecker's delta.
We extend this algebra to be $\mcal{A}$ containing nice functions of these generators
as well as polynomials so that we can do calculation in $\mcal{A}$ like
\begin{equation}
	-i\left[p_{j}^{\mu},f\left(\{x_{j}^{\mu}\}\right)\right]=\frac{\del f}{\del x_{j}^{\mu}}\left(\{x_{j}^{\mu}\}\right)
\end{equation}
for a function $f$ of generators.
We assume $\mcal{A}$ is represented on a Hilbert space $\mcal{H}(\Lambda)$
that carries data of the box $\Lambda=[0,L]^{d}\subset \mbb{R}^{d}$ confining particles.
Although that our discussion on the algebra $\mcal{A}$ extends to that on the representation $\mcal{H}(\Lambda)$
has to be verified for each representation since any representation of $\mcal{A}$ involves unbounded operators,
we do not get hung up on this point
and assume any algebraic observation on $\mcal{A}$ also holds on $\mcal{H}(\Lambda)$.

Let us assume the form of the Hamiltonian as
\begin{equation}
  H_{U}(\kappa)=H_{0}+\kappa U.
\end{equation}
Here $H_{0}$ is the part without nonuniform potential given by
\begin{equation}
	H_{0}=\sum_{j=1}^{N}\sum_{\mu=1}^{d}\frac{(p_{j}^{\mu})^{2}}{2m} + \sum_{j<k}w(\vc{x}_{j}-\vc{x}_{k}),
\end{equation}
where we shortly write $\vc{x}_{j}=(x_{j}^{1},\cdots,x_{j}^{d})$,
and $U$ is nonuniform potential
\begin{equation}
	U=\sum_{j=1}^{N}u(\vc{x}_{j}).
\end{equation}
In the Hamiltonian $H_{U}(\kappa)$, a parameter $\kappa$ controls the strength of nonuniform potential.
It is noted that $H_{0}$ may contain interparticle interaction.
The subscript $0$ does not mean the Hamiltonian is one for free particles,
but means it is without nonuniform potential.
Here we assume the functions $w$ and $u$ are sufficiently smooth ones so that
the interaction and the nonuniform potential define elements of $\mcal{A}$,
and moreover they are represented by bounded operators.

In Sect. \ref{sec:introduction}, we introduced the concept of NCRI in the setting of liquid in a rotating container,
but now we move on to a situation in which we slide the walls of the container with small velocity $v$ to some direction.
We can fairly expect the equivalence of these two pictures
when we focus on NCRI, which captures the asymptotic behavior at $v\to 0$.
To realize this situation, we assume the periodic boundary conditions are imposed along the $d$-th axis on the representation $\mcal{H}(\Lambda)$,
and slide the walls of the container with the velocity $v$ along the same direction.
The state that realizes at zero temperature is described by the ground state $\Psi(\kappa,v)$ of the Hamiltonian
\begin{equation}
  H(\kappa,v)=H_{U}(\kappa)-v P_{d}.
\end{equation}
Here $P_{d}$ is the $d$-th component of the total momentum $\vc{P}=(P_{1},\cdots, P_{d})$ with
\begin{equation}
	P_{\mu}=\sum_{j=1}^{N}p_{j}^{\mu},
\end{equation}
for $\mu=1,\cdots, d$.
Since $H_{0}$ does not contain nonuniform potential and translation invariant, we have $[H_{0},P_{d}]=0$.

Now we can see under a natural assumption that $H(\kappa,v)$ and $H(\kappa,-v)$ have the same spectrum.
Here the natural assumption is on the existence of an antilinear antiunitary involution $K$ such that
$\{K,P_{d}\}:=KP_{d}+P_{d}K=0$ and $[K,H_{U}(\kappa)]=0$.
Then for an eigenvector $\Psi$ of $H(\kappa,v)$, $K\Psi$ is an eigenvector of $H(\kappa,-v)$ corresponding to the same eigenvalue.
The operator $K$ is usually realized as complex conjugation.

Let us assume the ground state of $H_{U}(\kappa)$ is unique,
then the ground state energy $E(\kappa,v)$ of  $H(\kappa,v)$ is an analytic function of $v$ near $v=0$ and thus is expanded in powers of $v$ as
\begin{equation}
  E(\kappa,v)=E_{U}(\kappa)+c_{2}(\kappa)v^2+ O(v^4).
\end{equation}
Terms with odd powers of $v$ do not appear, since $E(\kappa, v)$ is an even  function of $v$.
We note that our assumption of uniqueness of the ground state is proved for spinless Bose systems in general situation.

The superfluid fraction $\rho_{\rm{s}}(\kappa)/\rho$ is defined in terms of the coefficient $c_{2}(\kappa)$ of $v^{2}$.
We introduce the effective mass $M_{\mrm{eff}}$ by $\braket{\Psi(\kappa,v),P_{d}\Psi(\kappa,v)}=M_{\mrm{eff}}v+O(v^{2})$,
which is interpreted as the mass of component flowing with the moving walls for sufficiently small $v$.
By using the Hellmann-Feynman theorem we can verify that $M_{\mrm{eff}}=-\frac{\del^{2}}{\del v^{2}}E(\kappa,v)\big|_{v=0}=-2c_{2}(\kappa)$.
Then we define the superfluid fraction as the mass fraction of stationary component by
\begin{equation}
  \label{eq:def_rs}
  \frac{\rho_{\rm{s}}(\kappa)}{\rho}=\frac{Nm-M_{\mrm{eff}}}{Nm}=1+\frac{2}{Nm}c_{2}(\kappa).
\end{equation}
This is an essential quantity for superfluidity characterizing NCRI of the system.
If the liquid is completely stationary while the walls are sliding along the $d$-th axis, we have $\rho_{\rm{s}}/\rho=1$,
and if the total liquid moves with the walls, we have $\rho_{\rm{s}}/\rho=0$.

We can see that $c_{2}(0)=0$ and thus $\rho_{\rm{s}}(0)/\rho=1$,
since $[H_{U}(0),P_{d}]=0$.
This fact implies physically that
without nonuniform potential, any liquid does not move with the sliding walls,
since the sliding walls are the same as stationary ones.
The difference of superfluid fraction from unity is due to the presence of nonuniform potential.
We are then led to ask the behavior of $c_{2}(\kappa)$ to understand superfluidity.

We close this section by making a remark on when the thermodynamic limit is taken.
In the above definition of the superfluid fraction, we first expand the ground state energy in terms of $v$ for a finite system,
and then take the thermodynamic limit so that the coefficient $c_{2}(\kappa)$ defines the superfluid fraction.
There is another way, in which one first takes the thermodynamic limit keeping $v\neq 0$,
and makes expansion of the ground state energy in terms of $v$ in which the coefficient of $v^{2}$ also defines the superfluid fraction.
In the second manner, it is possible to observe the superfluid fraction less than unity even without nonuniform potential,
as is easily checked for noninteracting systems.
Both of these two approaches are widely taken in study of superfluidity of the ground state
and which one is correct has not gained consensus.
We choose ours, which is more convenient for our purpose to study superfluidity of interacting systems as well as noninteracting ones,
since treating the term $-vP_{d}$ unperturbatively is extremely difficult for interacting systems except for some special cases.

\section{Construction of perturbation theory}
\label{sec:perturbation}
In the previous section, we introduced the superfluid fraction, which is fundamental quantity for the study of superfluidity.
It clearly depends on the strength of nonuniform potential $\kappa$,
then our interest is in this dependence.
In this section, we establish the perturbation expansion of the superfluid fraction in terms of $\kappa$.

In our Hamiltonian $H(\kappa, v)$,
the term $-vP_{d}$ is regarded as a perturbation to $H_{U}(\kappa)$.
Then $c_{2}(\kappa)$ is nothing but the perturbation coefficient of second order
and calculated in the standard way.
Here our aim is to investigate the dependence of $c_{2}(\kappa)$ on $\kappa$,
especially expand it in terms of $\kappa$.
Thus we are seemingly required to expand all excited states of $H_{U}(\kappa)$, which is not easy.
To avoid this difficulty, we write $c_{2}(\kappa)$ as the following integral
\begin{equation}
	\label{eq:coefficient2}
	c_{2}(\kappa)=\frac{1}{2\pi i}\int_{K_{\epsilon}}\frac{\braket{\Omega (\kappa),P_{d}(H_{U}(\kappa)-E)^{-1}P_{d}\Omega (\kappa)}}{E_{U}(\kappa)-E}dE,
\end{equation}
where $\Omega (\kappa)$ is the ground state of $H_{U}(\kappa)$ corresponding to the eigenvalue $E_{U}(\kappa)$.
$K_{\epsilon}=\{z\in\mbb{C}||z-E_{U}(\kappa)|=\epsilon\}\subset \mathbb{C}$, with $\epsilon$ being half of the first excitation energy of $H_{U}(\kappa)$,
is the closed integral contour such that $E_{U}(\kappa)$ is in the area enclosed by $K_{\epsilon}$ and 
other eigenvalues are not.
We remark the expression like in Eq. (\ref{eq:coefficient2}) is found in a textbook on the analytic perturbation theory~\cite{Kato1995},
which is developed as a rigorous theory for perturbation of operators.

Then we can expand $c_{2}(\kappa)$ in powers of $\kappa$.
Let $\Omega_{0}$ be the ground state of $H_{0}$ corresponding to the ground state energy $E_{0}$.
By definition of the operator norm, we have
$\|U\Psi\| \le \|U\|\|\Psi\|$
for an arbitrary $\Psi\in \mcal{H}(\Lambda)$,
implying that $U$ is relatively bounded with respect to $H_{0}$
Thus we can use the analytic perturbation theory to obtain $\Omega(\kappa)$ and $E_{U}(\kappa)$ for
$\kappa\in I_{U}(\Lambda):=(-(E_{1}-E_{0})/4\|U\|,(E_{1}-E_{0})/4\|U\|)$,
with $E_{1}-E_{0}$ being the first excitation energy of $H_{0}$.
The eigenstate $\Omega (\kappa)$ of $H_{U}(\kappa)$ is given by
\begin{equation}
	\label{eq:ground_state}
  	\Omega(\kappa)=Q(\kappa)\Omega_{0},
\end{equation}
where $Q(\kappa)$ is defined for $\kappa\in I_{U}(\Lambda)$ and is expanded as
\begin{align}
  Q(\kappa)&=\sum_{n=0}^{\infty}\kappa^{n}Q_{n}, \\
    \label{eq:Q}
  Q_{n}&=-\frac{1}{2\pi i}\int_{K_{\pr{\epsilon}}}(-1)^{n}T^{n}_{H_{0},U}dE,\\
  T^{n}_{H_{0},U}&:=(H_{0}-E)^{-1}[U(H_{0}-E)^{-1}]^{n}.
\end{align}
Here the integral contour goes counterclockwise along
$K_{\pr{\epsilon}}=\{z\in \mathbb{C}||z-E_{0}|=\pr{\epsilon}\}$
with $\pr{\epsilon}=(E_{1}-E_{0})/2$.
Since $K_{\pr{\epsilon}}$ is a subset in the resolvent set of $H_{0}$,
the integrand in Eq. (\ref{eq:Q}) is regarded as a Banach space valued function,
and thus the integral is understandable.
The ground state energy $E_{U}(\kappa)$ of $H_{U}(\kappa)$ is also analytic with respect to $\kappa\in I_{U}(\Lambda)$ and obtained as
\begin{equation}
	\label{eq:ground_energy}
  E_{U}(\kappa)=E_{0}+\kappa\frac{\braket{U\Omega_{0},Q(\kappa)\Omega_{0}}}{\braket{\Omega_{0},Q(\kappa)\Omega_{0}}}.
\end{equation}

We further prepare a useful expression for the resolvent $(H_{U}(\kappa)-E)^{-1}=(H_{0}+\kappa U-E)^{-1}$
in Eq. (\ref{eq:coefficient2}).
First we note that $H_{0}+\kappa U-E$ is also written as $H_{0}+\kappa U-E=[I+\kappa U(H_{0}-E)^{-1}](H_{0}-E)$.
Now we assume $E$ is in resolvent set of $H_{0}$, thus the resolvent $(H_{0}-E)^{-1}$ is well-defined.
It is well known that
if $\|\kappa U(H_{0}-E)^{-1}\| <1$, then $I+\kappa U(H_{0}-E)^{-1}$ is bijective and its inverse operator is bounded,
which is given by the following Neumann series
\begin{equation}
  [I+\kappa U(H_{0}-E)^{-1}]^{-1}=\sum_{n=0}^{\infty}(-1)^{n}\kappa^{n}[U(H_{0}-E)^{-1}]^{n}.
\end{equation}
Let us add further discussion on the condition $\|\kappa U(H_{0}-E)^{-1}\|<1$.
From a property of the operator norm, we have $\|\kappa U(H_{0}-E)^{-1}\|\le |\kappa| \|U\|\|(H_{0}-E)^{-1}\|$.
It is known that
\begin{equation}
  \|(H_{0}-E)^{-1}\|=\left(\inf_{s\in\sigma(H_{0})}|s-E|\right)^{-1},
\end{equation}
with $\sigma (H_{0})$ denoting the spectrum of $H_{0}$.
If $E\in K_{\pr{\epsilon}}$, then we have $\|(H_{0}-E)^{-1}\| = 2(E_{1}-E_{0})^{-1}$.
Thus a sufficient condition for $\|\kappa U(H_{0}-E)^{-1}\|<1$ is that $\kappa\in I_{U}(\Lambda)$.
Finally we obtain a series expression
\begin{equation}
	\label{eq:resolvent}
	(H_{0}(\kappa)-E)^{-1}=\sum_{n=0}^{\infty}(-1)^{n}\kappa^{n}(H_{0}-E)^{-1}[U(H_{0}-E)^{-1}]^{n}
\end{equation}
for $\kappa\in I_{U}(\Lambda)$.

Our next task is to substitute Eqs. (\ref{eq:ground_state}), (\ref{eq:ground_energy}) and (\ref{eq:resolvent}) 
into Eq. (\ref{eq:coefficient2}) and rearrange the integrand into a power series of $\kappa$.
Then we obtain
\begin{equation}
	\label{eq:coefficient2_series}
  c_{2}(\kappa)=\frac{1}{2\pi i}\int_{K_{\epsilon}}\left(\sum_{n=0}^{\infty}\kappa^{n}r_{n}(E)\right) dE.
\end{equation}
Here $r_{n}(E)$ are given by
\begin{align}
  & r_{0}(E)=\frac{g_{0}(E)}{E_{0}-E},\\
  & r_{n}(E)=\frac{1}{E_{0}-E}\left(g_{n}(E)-\sum_{k=0}^{n-1}f_{n-k}r_{k}(E)\right)\ \ (n\ge 1),
\end{align}
with
\begin{align}
  &f_{1}=a_{1} \hspace{15pt} f_{n}=a_{n}-\sum_{k=1}^{n-1}b_{n-k}f_{k}\ \ (n\ge 2), \\
  &g_{n}(E)=\sum_{i=0}^{n}\sum_{j=0}^{n-i}(-1)^{j}\braket{Q_{i}\Omega_{0},P_{d}T^{n}_{H_{0},U}P_{d}Q_{n-i-j}\Omega_{0}}, \\
  &a_{n}=\braket{U\Omega_{0},Q_{n-1}\Omega_{0}}\ \ (n\ge 1)\\
  &b_{n}=\braket{\Omega_{0},Q_{n}\Omega_{0}}\ \ (n\ge 0).
\end{align}

\begin{figure}
  \begin{center}
    \includegraphics[width=7cm]{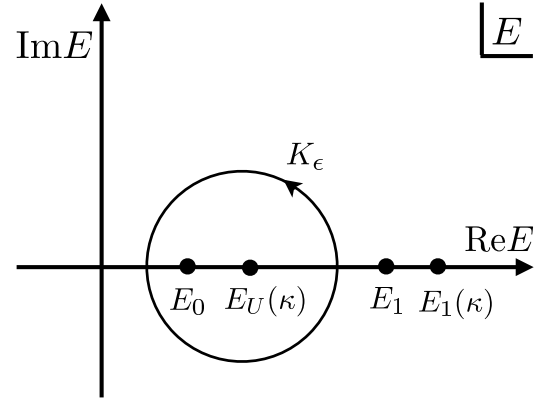}
    \caption{Integral contour in the complex plane. $E_{0}$ and $E_{1}$ denote the ground and the first excited state energy of $H_{0}$, respectively, and $E_{U}(\kappa)$ and $E_{1}(\kappa)$ denote the ground and first excited state energy of $H_{U}(\kappa)$, respectively.}
    \label{fig:integral_contour}
  \end{center}
\end{figure}

Here we make two assumptions.
First one is that we can commute the integral and infinite series to obtain
\begin{equation}
	\label{eq:c2series}
	c_{2}(\kappa)=\sum_{n=0}^{\infty}\kappa^{n}\frac{1}{2\pi i}\int_{K_{\epsilon}}r_{n}(E) dE.
\end{equation}
Secondly we assume, as shown in Fig. \ref{fig:integral_contour}, the integral contour $K_{\epsilon}$ surrounds $E_{0}$ as well as $E_{U}(\kappa)$,
to make our following computation possible.

The integrands in coefficients of $\kappa^{0}$ and $\kappa^{1}$ vanish as verified in Appendix\ref{sec:appendixa},
thus the leading contribution comes from the term of $\kappa^{2}$,
which is computed as follows.
$r_{2}(E)$ is
\begin{equation}
  r_{2}(E)=\frac{1}{E_{0}-E}\left(g_{2}(E)-\sum_{i=0}^{1}f_{1-k}r_{k}(E)\right)=\frac{g_{2}(E)}{E_{0}-E}.
\end{equation}
Here we used in the second equality $r_{0}(E)=r_{1}(E)=0$, which is shown in Appendix\ref{sec:appendixa}.
$g_{2}(E)$ is calculated as
\begin{equation}
  g_{2}(E)=\sum_{i=0}^{2}\sum_{j=0}^{2-i}(-1)^{j}\braket{Q_{i}\Omega_{0},P_{d}T^{j}_{H_{0},U}P_{d}Q_{2-i-j}\Omega_{0}}.
\end{equation}
The only nonvanishing term in this summation is the term of $i=1$ and $j=0$,
since otherwise $Q_{i}$ or $Q_{2-i-j}$ has to be $Q_{0}$,
the projection onto the one dimensional subspace spanned by $\Omega_{0}$ satisfying $P_{d}\Omega_{0}=0$,
and the matrix element vanishes.
Since $H_{0}$ is self-adjoint, $Q_{1}$ is also self-adjoint.
Thus we get
\begin{equation}
  g_{2}(E)=\braket{\Omega_{0}, Q_{1}P_{d}(H_{0}-E)^{-1}P_{d}Q_{1}\Omega_{0}}.
\end{equation}
Substituting the form of $Q_{1}$ in Eq (\ref{eq:Q}) for $n=1$
and using $(H_{0}-E)^{-1}\Omega_{0}=(E_{0}-E)^{-1}\Omega_{0}$, we can calculate $g_{2}(E)$.
We take as a complete orthonormal basis of $\mcal{H}(\Lambda)$ 
the set of eigenvectors $\{\Omega_{n}\}_{n}$ of $H_{0}$ satisfying $H_{0}\Omega_{n}=E_{n}\Omega_{n}$.
Then the integral along $K_{\pr{\epsilon}}$ is easily computed leading us to
\begin{equation}
  g_{2}(E)=\sum_{n=1}^{\infty}\frac{|\braket{\Omega_{n},[U,P_{d}]\Omega_{0}}|^{2}}{(E_{n}-E_{0})^{2}(E_{n}-E)}.	
\end{equation}
Since $\braket{\Omega_{0},[U,P_{d}]\Omega_{0}}=0$, the summation in the above equation runs from $n=1$.
The coefficient of $\kappa^{2}$ is computed as
\begin{equation}
  \frac{1}{2\pi i}\int_{K_{\epsilon}} \frac{g_{2}(E)}{E_{0}-E} dE
  	=-\sum_{n=1}^{\infty}\frac{|\braket{\Omega_{n},[U,P_{d}]\Omega_{0}}|^{2}}{(E_{n}-E_{0})^{3}}.
\end{equation}

We then obtain the perturbation expansion of $c_{2}(\kappa)$ as follows:
\begin{equation}
  c_{2}(\kappa)=-\kappa^{2}\sum_{n=1}^{\infty}\frac{|\braket{\Omega_{n},[U,P_{d}]\Omega_{0}}|^{2}}{(E_{n}-E_{0})^{3}}+O(\kappa^{3}).
\end{equation}

Here we rewrite the commutator $[U, P_{d}]$ in a more convenient form.
In our algebra $\mcal{A}$ given by Eq. (\ref{eq:CCR}),
the adjoint action of $-ip_{j}^{\mu}$ on a polynomial of $\{x_{j}^{\mu}\}$
differentiates the polynomial by $x_{j}^{\mu}$.
We assume this observation is valid even if the polynomial above is replaced by a more general function of $\{x_{j}^{\mu}\}$ in $\mcal{A}$.
Thus we can see the commutator $[U,P_{d}]$ is a function only of $\{x_{j}^{\mu}\}$
\begin{equation}
  [U,P_{d}]=-i U_{d},
\end{equation}
where $U_{d}$ is
\begin{equation}
	U_{d}=\sum_{j=1}^{N}u_{d}(\vc{x}_{j}),
\end{equation}
with $u_{d}$ standing for $\frac{\del u}{\del x^{d}}$.
Then the expansion of the superfluid fraction becomes 
\begin{equation}
	\label{eq:expansion}
  \frac{\rho_{\rm{s}}}{\rho}(\kappa)
  	=1-\frac{2\kappa^2}{Nm}\sum_{n=1}^{\infty}\frac{|\braket{\Omega_{n},U_{d}\Omega_{0}}|^2}{(E_{n}-E_{0})^{3}}+O(\kappa^3).
\end{equation}

It will be convenient to express the coefficient of $\kappa^{2}$ using the dynamic structure factor.
$\{\Omega_{n}\}_{n}$ are eigenvectors of the Hamiltonian which commutes with the momentum.
Thus we can choose $\{\Omega_{n}\}_{n}$ such that each $\Omega_{n}$ is also an eigenvector of the momentum,
and the set of eigenvectors of the Hamiltonian chosen in this way can be denoted by $\{\Omega_{\vc{k},n}\}_{\vc{k},n}$.
Each $\Omega_{\vc{k},n}$ is a simultaneous eigenvector so that
$H_{0}\Omega_{\vc{k},n}=E_{\vc{k},n}\Omega_{\vc{k},n}$ and $\vc{P}\Omega_{\vc{k},n}=\vc{k}\Omega_{\vc{k},n}$.
The ground state is denoted by $\Omega_{\vc{0},0}$, 
since it is an eigenvector of the momentum corresponding to $\vc{0}$ and it is the lowest energy state.

It is convenient to write $U_{d}$ in the following form:
\begin{equation}
  U_{d}=\int_{\Lambda} d\vc{x} u_{d}(\vc{x})\rho (\vc{x})=\frac{1}{V}\sum_{\vc{k}}ik_{d}\tilde{u}(\vc{k})\rho_{-\vc{k}},
\end{equation}
where $\rho(\vc{x})$ is the particle density operator formally written as $\rho(\vc{x})=\sum_{j=1}^{N}\delta (\vc{x}-\vc{x}_{j})$
and its Fourier transform $\rho_{\vc{k}}$ is defined by
\begin{equation}
  \rho_{\vc{k}}:=\int_{\Lambda} \rho(\vc{x})e^{-i\vc{k}\cdot\vc{x}}d\vc{x}=\sum_{j=1}^{N}e^{-i\vc{k}\cdot\vc{x}_{j}}
\end{equation}
for $\vc{k}\in\frac{2\pi}{L}\mathbb{Z}^{d}$.
$V=L^{d}$ is the volume of the box $\Lambda=[0,L]^{d}$.
$\tilde{u}$ is the Fourier transform of the one-body potential $u$ defined in the same convention.

From the commutation relation $[\vc{P},\rho_{-\vc{k}}]=\vc{k}\rho_{-\vc{k}}$,
we see $\rho_{-\vc{k}}\Omega_{\vc{0},0}$ is an eigenvector of $\vc{P}$ corresponding to an eigenvalue $\vc{k}$,
and thus orthogonal to any vectors corresponding to differnt momentum.
This observation allows us to rewrite Eq. (\ref{eq:expansion}) into
\begin{equation}
	\label{eq:expansion_momentum}
  \frac{\rho_{\rm{s}}}{\rho}
	=1-\frac{2\kappa^{2}}{Nm}\sum_{(\vc{k},n)\neq (\vc{0},0)}\left(\frac{k_{d}|\tilde{u}(\vc{k})|}{V}\right)^{2}\frac{|\braket{\Omega_{\vc{k},n},\rho_{-\vc{k}}\Omega_{\vc{0},0}}|^{2}}{(E_{\vc{k},n}-E_{\vc{0},0})^{3}}+O(\kappa^{3}).
\end{equation}

Setting $\omega_{\vc{k},n}=E_{\vc{k},n}-E_{\vc{0},0}$,
the dynamic structure factor is given by
\begin{equation}
  S(\vc{k},\omega)=\sum_{n}|\braket{\Omega_{\vc{k},n},\rho_{-\vc{k}}\Omega_{\vc{0},0}}|^{2}\delta(\omega-\omega_{\vc{k},n}).
\end{equation}
Then the superfluid fraction is written as
\begin{equation}
	\label{eq:expansion_s}
  \frac{\rho_{\rm{s}}(\kappa)}{\rho}
  	=1-\frac{2\kappa^{2}}{Nm}\sum_{\vc{k}}\left(\frac{k_{d}|\tilde{u}(\vc{k})|}{V}\right)^{2}\int_{0}^{\infty} d\omega \frac{S(\vc{k},\omega)}{\omega^{3}}+O(\kappa^{3}).
\end{equation}

For later convenience, let us write down the expansion in case that the nonuniform potential is periodic $u(\vc{x})=u\cos (\vc{p}\cdot \vc{x})$,
with $\vc{p}\in \frac{2\pi}{L}\mbb{Z}^{d}$.
Assuming $S(\vc{k},\omega)=S(-\vc{k},\omega)$, we obtain
\begin{equation}
	\label{eq:expansion_periodic}
	\frac{\rho_{\mrm{s}}(\kappa)}{\rho}=1-\frac{(p_{d}u\kappa)^{2}}{Nm}\int_{0}^{\infty}d\omega \frac{S(\vc{p},\omega)}{\omega^{3}} +O(\kappa^{3}).
\end{equation}

We can also treat nonuniform potential as random potential.
Assume the one particle potential $u(\vc{x})$ is Gaussian random potential satisfying
\begin{align}
	\label{eq:ran_mean}
	\braket{u(\vc{x})}_{\mrm{r.a.}}&=0,\\
	\label{eq:ran_cov}
	\braket{u(\vc{x})u(\vc{y})}_{\mrm{r.a.}}&=\gamma^{2}\delta(\vc{x}-\vc{y}).
\end{align}
Here $\braket{\cdot}_{\mrm{r.a.}}$ denotes the average over randomness.
Then the Fourier coefficients satisfy
\begin{equation}
	\braket{\tilde{u}(\vc{k})\tilde{u}(\pr{\vc{k}})}_{\mrm{r.a.}}=\gamma^{2}V\delta_{\vc{k}+\pr{\vc{k}},0}.
\end{equation}
Averaging the both sides of Eq. (\ref{eq:expansion_s}) over randomness
with noticing $\tilde{u}(\vc{k})^{\ast}=\tilde{u}(-\vc{k})$, we have
\begin{equation}
	\label{eq:expansion_random}
	\Braket{\frac{\rho_{\mrm{s}}(\kappa)}{\rho}}_{\mrm{r.a.}}
		=1-\frac{2(\gamma\kappa)^{2}}{VNm}\sum_{\vc{k}}k_{d}^{2}\int_{0}^{\infty}d\omega \frac{S(\vc{k},\omega)}{\omega^{3}}+O(\kappa^{3}).
\end{equation}

Let us make a remark that the perturbation expansion of the superfluid fraction
is proved to converge only for finite systems.
The convergence radius is expected to get $0$ at the thermodynamic limit, {\it i.e.} $\bigcap_{\Lambda}I_{U}(\Lambda)=\{0\}$, because of two reasons.
Firstly the first excitation gap closes at the thermodynamic limit in many examples.
Secondly the operator norm of nonuniform potential diverges at the thermodynamic limit since it is expected to be proportional to $N$.
Thus contributions from the terms of higher order are not assured to be small relative to that from the leading term at the thermodynamic limit.
In spite of this mathematical difficulty, we formally take the thermodynamic limit of the expansion
and try to discuss superfluidity by investigating behavior of the leading coefficient.
If the coefficient of $\kappa^{2}$ converges, we can interpret the system exhibits NCRI
since the superfluid fraction approaches unity for sufficiently small $\kappa$.
In this paper, we assume the convergence of the coefficient of $\kappa^{2}$ gives a sufficient condition to NCRI.
The coefficient may diverge and in this case, we have to be careful to make physical interpretation.
One interpretation of the divergent coefficient is that the superfluidity breaks down
under infinitesimally small nonuniform potential at the thermodynamic limit.
The other one is that the $\kappa$-dependence of the superfluid fraction
near $\kappa=0$ is not quadratic at the thermodynamic limit.
We cannot conclude which interpretation is correct in our formulation,
and thus in such a case nonperturbative treatment of nonuniform potential is required.

As seeing the perturbation expansion in Eq.(\ref{eq:expansion_s}),
we find that large density fluctuation at low energy suppresses the superfluid fraction.
Especially divergence of the coefficient is due to the singularity of the integrand at $\omega=0$.
In case of periodic potential oscillating with wave number $\vc{p}$,
the coefficient in Eq. (\ref{eq:expansion_periodic}) is expected to converge at the thermodynamic limit
if the excitation spectrum has a gap at the momentum $\vc{p}$,
since the integrand does not have singularity at low energy.

\section{Bounds for superfluid fraction}
\label{sec:bounds}
In the previous section, we derived the perturbation expansion of the superfluid fraction 
in terms of the strength of nonuniform potential.
We also saw that the coefficient of leading order is written using the dynamic structure factor.

In this section, we estimate upper and lower bounds of the superfluid fraction
perturbatively up to the leading order.

\subsection{Upper Bound}

We derive an upper bound by estimating the integral in Eq. (\ref{eq:expansion_s}).
Let us introduce Jensen's inequailty, which says
\begin{prop}[Jensen's inequality~\cite{LL2001}]
Let $I\subset \mbb{R}$ be a connected subset and
$p:I\to \mbb{R}$ be a probability distribution function, {\it i.e.},
a function satisfying $p(x)\ge 0$ for all $x\in I$ and $\int_{I}p(x)dx=1$.
We assume $\phi:I\to\mathbb{R}$ is a convex function, {\it i.e.}, it satisfies
\begin{equation}
  \phi(tx+(1-t)y) \le t\phi(x)+(1-t)\phi(y)
\end{equation}
for arbitrary $x, y\in I$ and $t\in [0,1]$.
Then we have
\begin{equation}
  \int_{I}\phi(x)p(x)dx \ge \phi\left(\int_{I}xp(x)dx\right),
\end{equation}
if the both sides exist.
\end{prop}

Let us introduce the static structure factor related to the dynamic one by
\begin{equation}
	\label{eq:static_structure_factor}
	\tilde{S}(\vc{k})=\frac{1}{N}\int_{0}^{\infty}d\omega S(\vc{k},\omega)=\frac{1}{N}\sum_{n}|\braket{\Omega_{\vc{k},n},\rho_{-\vc{k}}\Omega_{\vc{0},0}}|^{2}.
\end{equation}
Then a function $p_{\vc{k}}(\omega)=S(\vc{k},\omega)/N\tilde{S}(\vc{k})$,
defines a probability distribution function on $I=\mbb{R}_{>0}$.
Thus we can apply Jensen's inequality to the integral in Eq. (\ref{eq:expansion_s}) to obtain
\begin{equation}
	\int_{0}^{\infty}d\omega \frac{S(\vc{k},\omega)}{\omega^{3}}
		\ge \frac{(N\tilde{S}(\vc{k}))^{4}}{\left(\int_{0}^{\infty}d\omega\omega S(\vc{k},\omega)\right)^{3}}
		=\frac{8m^{2}N}{|\vc{k}|^{6}}\tilde{S}(\vc{k})^{4}
\end{equation}
since $1/\omega^{3}$ is convex on $I$.
Here we used the $f$-sum rule
\begin{equation}
	\int_{0}^{\infty}d\omega \omega S(\vc{k},\omega)=\frac{N|\vc{k}|^{2}}{2m}
\end{equation}
in the equality.
Consequently we obtain an upper bound for the superfluid fraction
\begin{equation}
	\label{eq:upper_bound}
	\frac{\rho_{\mrm{s}}(\kappa)}{\rho}
		\le 1-(4m\kappa)^{2}\sum_{\vc{k}}\left(\frac{k_{d}|\tilde{u}(\vc{k})|}{V|\vc{k}|^{3}}\right)^{2}\tilde{S}(\vc{k})^{4} +O(\kappa^{3}).
\end{equation}

For periodic potential $u(\vc{x})=u\cos(\vc{p}\cdot\vc{x})$, the upper bound becomes
\begin{equation}
	\label{eq:upper_bound_periodic}
	\frac{\rho_{\mrm{s}}(\kappa)}{\rho}\le 1-2\left(\frac{2p_{d}mu\kappa}{|\vc{p}|^{3}}\right)^{2}\tilde{S}(\vc{p})^{4}+O(\kappa^{3}).
\end{equation}

When the nonuniform potential is random potential defined by Eqs. (\ref{eq:ran_mean}) and (\ref{eq:ran_cov}),
we take the random average of Eq. (\ref{eq:upper_bound}) to obtain
\begin{equation}
	\label{eq:upper_bound_random}
	\Braket{\frac{\rho_{\mrm{s}}(\kappa)}{\rho}}_{\mrm{r.a.}}\le 1-\frac{(4m\gamma\kappa)^{2}}{V}\sum_{\vc{k}}\frac{k_{d}^{2}\tilde{S}(\vc{k})^{4}}{|\vc{k}|^{6}}+O(\kappa^{3}).
\end{equation}
Although a similar upper bound for the superfluid fraction under random potential
is also obtained by Kim {\it et al.}~\cite{KS1993},
our bound in Eq. (\ref{eq:upper_bound_random}) is more general
since it does not assume the existence of momentum cutoff,
which is imposed on the random potential in their analysis.

\subsection{Lower Bound}
Let us derive a lower bound for superfluid fraction starting from Eq. (\ref{eq:expansion_momentum}).
We set
\begin{equation}
  \Delta(\tilde{u}):=\inf_{n,\vc{k}\in\mathrm{supp}(\tilde{u})}(E_{\vc{k},n}-E_{\vc{0},0}),
\end{equation}
and we get a lower bound of the superfluid fraction as
\begin{align}
  \frac{\rho_{\rm{s}}}{\rho}
  	&\ge 1-\frac{2\kappa^{2}}{Nm}\sum_{\vc{k},n}\left(\frac{k_{d}|\tilde{u}(\vc{k})|}{V}\right)^{2}\frac{|\braket{\Omega_{\vc{k},n},\rho_{-\vc{k}}\Omega_{\vc{0},0}}|^{2}}{\Delta(\tilde{u})^{3}}+O(\kappa^{3}) \notag \\
  	\label{eq:lower_bound}
  	&=1-\frac{2\kappa^{2}}{m \Delta(\tilde{u})^{3}}\sum_{\vc{k}}\left(\frac{k_{d}|\tilde{u}(\vc{k})|}{V}\right)^{2}\tilde{S}(\vc{k}) +O(\kappa^{3}).
\end{align}
Here we used the form of the static structure factor in Eq. (\ref{eq:static_structure_factor}).
We see that a lower bound of the superfluid fraction is expressed by the static structure factor and data of the spectral infimum.
We again recognize that if $\Delta(\tilde{u})>0$, {\it i.e.},
there exists a gap in spectrum for wave number $\vc{k}$ in the support of $\tilde{u}$,
the superfluid fraction tends to unity for sufficiently weak nonuniformity as $\kappa \to 0$.

For a periodic potential $u(\vc{x})=u\cos (\vc{p}\cdot\vc{x})$, we replace $\Delta(\tilde{u})$
by the excitation gap at momentum $\vc{p}$, $\Delta(\vc{p}):=\inf_{n}(E_{\vc{k},n}-E_{\vc{0},0})$ to obtain
\begin{equation}
	\label{eq:lower_bound_periodic}
	\frac{\rho_{\mrm{s}}(\kappa)}{\rho}\ge 1-\frac{(k_{d}u\kappa)^{2}}{m\Delta(\vc{p})^{3}}S(\vc{p})+O(\kappa^{3}).
\end{equation}

When the nonuniform potential is realized by random potential defined by Eqs. (\ref{eq:ran_mean}) and (\ref{eq:ran_cov}),
we replace $\Delta(\tilde{u})$ by the first excitation energy $E_{1}-E_{0}$
and take random average to obtain a lower bound
\begin{equation}
	\Braket{\frac{\rho_{\mrm{s}}(\kappa)}{\rho}}_{\mrm{r.a.}}
		\ge 1-\frac{2(\gamma\kappa)^{2}}{m(E_{1}-E_{0})^{3}}\frac{1}{V}\sum_{\vc{k}}k_{d}^{2}\tilde{S}(\vc{k})+O(\kappa^{3}).
\end{equation}
This lower bound under random potential is useless if the first excitation energy
approaches $0$ as taking $L\to \infty$,
but still may give a meaningful lower bound for a fully gapped system.

\section{Examples}
\label{sec:examples}
Following the results obtained in the previous sections,
we investigate superfluid properties in several examples
fixing representations $\mcal{H}(\Lambda)$ concretely.

\subsection{Ideal Bosons}
In this and next subsections we take the space of square integrable functions in a box $\Lambda=[0,L]^{d}$ with periodic boundary conditions
as the one particle Hilbert space $\mfrak{h}$,
and let $\mcal{H}(\Lambda)$ be the $N$-fold symmetrized tensor product of $\mfrak{h}$.

We have the exact dynamic structure factor of a noninteracting Bose system as
\begin{equation}
  S(\vc{k},\omega)=N \delta(\omega-\omega_{\vc{k}}),
\end{equation}
where $\omega_{\vc{k}}=\frac{|\vc{k}|^{2}}{2m}$ is the one-particle excitation spectrum.
Thus the superfluid fraction is obtained as
\begin{equation}
  \frac{\rho_{\rm{s}}}{\rho}
  	=1-16m^{2}\kappa^{2}\sum_{\vc{k}}\left(\frac{k_{d}|\tilde{u}(\vc{k})|}{V|\vc{k}|^{3}}\right)^{2}+O(\kappa^{3}).
\end{equation}

Let us consider the case that the one-particle potential is given by $u(\vc{x})=u\cos (\vc{p}\cdot\vc{x})$.
The superfluid fraction in this case Eq. (\ref{eq:expansion_periodic}) becomes
\begin{equation}
  \frac{\rho_{\rm{s}}}{\rho}=1-2\left(\frac{2mup_{d}}{|\vc{p}|^{3}}\right)^{2}\kappa^{2}+O(\kappa^{3}).
\end{equation}
We can see the coefficient is finite for any nonzero $\vc{p}$,
implying its superfluidity is stable under the periodic potential. 
As we take long wavelength limit $|\vc{p}|\to 0$ fixing the direction $\vc{p}/|\vc{p}|$,
the coefficient of $\kappa^{2}$ diverges to infinity due to its parabolic spectrum.
Reflecting this fact, the coefficient of $\kappa^{2}$ under Gaussian random potential in Eq. (\ref{eq:expansion_random}) also diverges.

Actually, without use of our perturbation theory,
the superfluid fraction of ideal bosons under periodic potential is calculated as
$m/m^{\ast}$ with $m^{\ast}$ being the effective mass at the zero wavenumber~\cite{Leggett1973}.
Thus we can say the divergence of coefficient is an expression of the breakdown of superfluidity
due to diverging $m^{\ast}$ under long wavelength potential.

\subsection{Interacting Bosons}
We obtained in the previous section a lower bound for the superfluid fraction in Eq. (\ref{eq:lower_bound})
and discussed that
if the excitation spectrum has a gap for $\vc{k}\in \supp(\tilde{u})$,
the superfluid fraction tends to $1$ as $\kappa\to 0$.
For many bose systems there are evidences that this condition is satisfied.
Typically the dynamic structure factor $S(\vc{k},\omega)$ has a strong peak
at $\omega=\epsilon(\vc{k})$.
Only taking this mode, we approximate the dynamic structure factor as
\begin{equation}
	\label{eq:single_mode_dynamic_structure}
	S(\vc{k},\omega)=\frac{N|\vc{k}|^{2}}{2m\epsilon(\vc{k})}\delta(\omega-\epsilon(\vc{k})).
\end{equation}
Here the prefactor of the delta function is determined such that
it satisfies the $f$-sum rule.

We consider a case that the potential is given by $u(\vc{x})=u\cos (\vc{p}\cdot \vc{x})$.
Using the approximated dynamic structure factor in Eq. (\ref{eq:single_mode_dynamic_structure}),
our perturbation expansion in Eq. (\ref{eq:expansion_periodic}) becomes
\begin{equation}
	\label{eq:expansion_interacting_bosons}
	\frac{\rho_{\mrm{s}}(\kappa)}{\rho}
		=1-2\left(\frac{p_{d}|\vc{p}|u\kappa}{2m\epsilon(\vc{p})}\right)^{2}+O(\kappa^{3}).
\end{equation}

The dynamic structure factor of the form in Eq. (\ref{eq:single_mode_dynamic_structure})
induces an approximation of the static structure factor as
\begin{equation}
	\label{eq:single_mode_static_structure}
  \tilde{S}(\vc{k}) \simeq \frac{|\vc{k}|^{2}}{2m\epsilon(\vc{k})}.
\end{equation}
Under this approximation, the lower bound for the superfluid fraction in Eq. (\ref{eq:lower_bound_periodic}) becomes
\begin{equation}
   \frac{\rho_{\rm{s}}}{\rho} \ge 1-2 \left(\frac{p_{d}|\vc{p}|u\kappa}{2m\epsilon(\vc{p})^{2}}\right)^{2}+O(\kappa^{3})
\end{equation}
with assuming $\Delta(\vc{p})=\epsilon(\vc{p})$.
We also find the upper bound in Eq. (\ref{eq:upper_bound_periodic}) becomes
\begin{equation}
  \frac{\rho_{\rm{s}}}{\rho} \le 1-2 \left(\frac{p_{d}|\vc{p}|u\kappa}{2m\epsilon(\vc{p})^{2}}\right)^{2}+O(\kappa^{3}).
\end{equation}
These lower and the upper bounds coincide with the expansion in Eq. (\ref{eq:expansion_interacting_bosons}),
implying our estimation is strict if the single mode approximated structure factor in Eqs.
(\ref{eq:single_mode_dynamic_structure}) and (\ref{eq:single_mode_static_structure}) well describe the density fluctuation of the system.

Let us investigate the case that the potential is a long wave length one.
We can fairly assume $\epsilon(\vc{p})$ depends linearly on $\vc{p}$ such as $\epsilon(\vc{p})\sim c|\vc{p}|$ as $|\vc{p}|\to 0$
with $c$ being the velocity of sound.
Then we obtain from Eq. (\ref{eq:expansion_interacting_bosons})
\begin{equation}
	\frac{\rho_{\mrm{s}}(\kappa)}{\rho}
		\sim 1-2\left(\frac{p_{d}u\kappa}{2mc|\vc{p}|}\right)^{2}+O(\kappa^{3})
\end{equation}
as $|\vc{p}|\to 0$.
We can see that the superfluidity of this system is,
differently from ideal boson,
stable against nonuniform potential of long wave length.

\subsection{Bose Gas in Mean Field Limit}
In this subsection we take the space of square integrable functions on the box $\Lambda=[0,1]^{d}$ with the periodic boundary conditions
as the one-particle Hilbert space $\mfrak{h}$ and let $\mcal{H}(\Lambda)$ be the $N$-fold symmetrized tensor product of $\mfrak{h}$.

The Hamiltonian to be focused on is given by
\begin{equation}
	H_{N}=-\frac{1}{2m}\sum_{j=1}^{N}\Delta_{j} + \frac{1}{N-1}\sum_{j<k}w(\vc{x}_{j}-\vc{x}_{k}).
\end{equation}
Remarkable points are that the system is confined in a fixed torus and the prefactor $1/(N-1)$ in front of the interaction term in the Hamiltonian.
Although this model seems to be artificial, Seiringer~\cite{Seiringer2011} proved that
the low energy excitation of this Hamiltonian admits the picture of Bogoliubov quasi-paticles at the mean-field limit $N\to \infty$.
Let us describe the situation more precisely.
We define the Fourier transform of the interaction potential as
\begin{equation}
	\tilde{w}(\vc{k}):=\int_{[0,1]^{d}}w(\vc{x})e^{-i\vc{k}\cdot\vc{x}}d\vc{x}
\end{equation}
for $\vc{k}\in (2\pi \mbb{Z})^{d}$.
Then from the Bogoliubov approximation we expect that excitation states are obtained by
exciting quasi-particles with momentum $\vc{k}$ and energy
\begin{equation}
	\epsilon(\vc{k})=\sqrt{\left(\frac{|\vc{k}|^{2}}{2m}\right)^{2}+\frac{|\vc{k}|^{2}}{m}\tilde{w}(\vc{k})}
\end{equation}
from the ground state.
This picture of excited states is actually valid for excited states of sufficiently low energy 
in systems with sufficiently many particles.
\begin{thm}[Seiringer~\cite{Seiringer2011}]
Let $E_{0}(N)$ be the ground state energy of $H_{N}$.
Then the spectrum $H_{N}-E_{0}(N)$ below an energy $\xi$ consists of finite sums of the form
\begin{equation}
	\sum_{\vc{k}\in(2\pi\mbb{Z})^{d}\backslash \{\vc{0}\}}\epsilon(\vc{k})n(\vc{k})+O\left(\xi^{3/2}N^{-1/2}\right),
\end{equation}
where $n(\vc{k})\in\mbb{Z}_{\ge 0}$.
Moreover, the excited state corresponding to $\{n(\vc{k})\}_{\vc{k}\in (2\pi\mbb{Z})^{d}}$ has total momentum $\sum_{\vc{k}}\vc{k}n(\vc{k})$.
\end{thm}

The above theorem allows us to expect the dynamic structure factor is well approximated by
the form given in Eq. (\ref{eq:single_mode_dynamic_structure}) 
for sufficiently small $\omega$ and sufficiently large $N$~\cite{PN1999}.
Then we use the dynamic structure factor approximated in this way to compute the perturbation coefficient
of the superfluid fraction.

Now we perturb this system by Gaussian random potential defined by equations (\ref{eq:ran_mean}) and (\ref{eq:ran_cov})
and compute superfluid fraction by equation (\ref{eq:expansion_random}).
For computational simplicity we assume $d=3$ and $w(\vc{x})=g\delta(\vc{x})$.
Then we have
\begin{equation}
	\Braket{\frac{\rho_{\mrm{s}}(\kappa)}{\rho}}_{\mrm{r.a.}}=1-\frac{m^{3/2}}{6\pi g^{1/2}}(\gamma\kappa)^{2}+O(\kappa^{3}).
\end{equation}
We remark in the above expression that stronger interaction leads to larger superfluid fraction,
with which a similar argument has already been made by K\"{o}nenberg {\it et al.}~\cite{KMSY2015} for one dimensional case

\subsection{Free Fermions}
In the following subsections we treat fermionic systems,
taking as $\mcal{H}(\Lambda)$ an antisymmetrized tensor product of space of square integrable functions.

The dynamic structure factor $S^{0}(\vc{k},\omega)$ for free fermions in three dimension is exactly calculated in textbooks~\cite{PN1999}
and its low frequency behavior for $|\vc{k}|<2k_{\rm{F}}$ is proportional to $\omega$ and $N$,
with $k_{\mrm{F}}$ being the Fermi wavenumber.
Thus the coefficient of $\kappa^{2}$ diverges to infinity at the thermodynamic limit.
Although divergence of the coefficient can be interpreted in several ways,
we can fairly interpret this divergence as implying that the superfluidity of free fermions is fragile under nonuniform potential, 
if there is $\vc{k}$ such that $|\vc{k}|<2k_{\rm{F}}$ and $\tilde{u}(\vc{k})\neq 0$.
The coefficient also diverges under random potential defined by equations (\ref{eq:ran_mean}) and (\ref{eq:ran_cov}).

Although we announced that we discuss superfluidity in the expansion up to the leading order,
let us make the following remark.
If $\tilde{u}$ has value only for $|\vc{k}|>2k_{\rm{F}}$, the coefficient of $\kappa^{2}$ converges.
However, this does not mean that a free fermi system exhibits superfluidity
if the nonuniform potential oscillates with large wave number $|\vc{k}|>2k_{\rm{F}}$.
Actually coefficients of higher order in the perturbation expansion is expected to diverge to infinity,
preventing free fermions from showing superfluidity.

\subsection{Interacting Fermions}

By calculation with the random phase approximation for fermions with Coulomb interaction,
we can see the dynamic structure factor satisfies~\cite{PN1999}
\begin{equation}
  S(\vc{k},\omega)\sim c(\vc{k})S^{0}(\vc{k},\omega),
\end{equation}
as $\omega\to 0$ with $c(\vc{k})$ depending only on $\vc{k}$.
The coefficient of $\kappa^{2}$ again diverges to infinity as in the free case.
This can also be interpreted that interacting fermions
does not exhibit superfluidity at the thermodynamic limit.

\subsection{Tomonaga-Luttinger Liquid}

Tomonaga-Luttinger liquid (TLL) describes low-energy states of
interacting fermions in one dimension \cite{Giamarchi2004}.
Its Hamiltonian is written after bosonization as
\begin{equation}
  H=\frac{c}{2\pi}\int dx \left[ K(\pi\Pi(x))^{2}+\frac{1}{K}(\nabla\phi(x))^{2}\right],
\end{equation}
where $\phi$ is a compactified Bose field and $\Pi$ is its conjugate momentum satisfying
\begin{equation}
  [\phi(x),\Pi(\pr{x})]=i\delta(x-\pr{x}).
\end{equation}
Two parameters $c$ and $K$ control TLL.
$c$ is interpreted as the renormalized Fermi velocity of the system.
Rather important is $K$, which is called the Luttinger parameter and characterizes the effective interaction between particles.
When $K>1$ (resp. $K<1$), the corresponding Luttinger liquid models
one-dimensional fermions with attractive (resp. repulsive) interaction,
which is said to possess superconducting (resp. charge density wave) state as the ground state.

In this subsection we investigate superfluidity of this system based on our perturbation expansion.
The excitation spectrum has gap at almost every wavenumber,
but it closes at integer times of $2k_{\rm{F}}$, with $k_{\mrm{F}}$ being the Fermi wavenumber.
Thus we have to carefully investigate the coefficient of $\kappa^{2}$
for periodic potential oscillating just by these wavenumbers.
Let us focus on the case of periodic potential oscillating with $k=2k_{\mrm{F}}$.
In Appendix\ref{sec:app_tll} we investigate the low energy behavior of the dynamic structure factor
at $k=2k_{\mrm{F}}$. It is
\begin{equation}
  S(k=2k_{\mrm{F}},\omega)\propto L\omega^{2(K-1)},
\end{equation}
with $L$ denoting the length of the system.
Thus the integral in the coefficient of $\kappa^{2}$
\begin{equation}
  \frac{1}{N}\int_{0}^{\infty}\frac{S(2k_{\rm{F}},\omega)}{\omega^{3}}d\omega
\end{equation}
is expected to converge for $K>2$.

Though TLL is said to exhibit superfluidity if $K>1$,
our result suggests that the superfluidity is robust against nonuniform potential if $K>2$.
We cannot make any conclusion on superfluidity of TLL for $1<K\le 2$ in our formulation,
since the coefficient diverges.
We need nonperturbative treatment of nonuniform potential
to discuss superfluidity of TLL corresponding to $1<K\le 2$
under nonuniform potential.

\subsection{Spinless chiral $p$-wave superfluid}
As mentioned before, if there exists energy gap in the excitation spectrum for every wave number,
the perturbation coefficient of $\kappa^{2}$ is expected to be finite at the thermodynamic limit,
and the superfluidity is robust against nonuniform potential.
From this observation, we can conclude that
the superfluidity of $s$-wave superfluid without spectral node is robust.
For nodal superfluid, however, it is nontrivial whether superfluidity is robust against
potential oscillating with the very wave number on which the particle-hole excitation spectrum become gapless.

In this subsection we investigate the superfluid property of spinless chiral $p$-wave (or an ABM state) superfluid in three dimension~\cite{Kita2015}.
The one-particle excitation spectrum is given by
\begin{equation}
  E_{\vc{k}}=\sqrt{\xi_{\vc{k}}^{2}+\Delta^{2}|\hat{k}_{\perp}|^{2}}.
\end{equation}
Here $\xi_{\vc{k}}$ is the spectrum of free particles measured from the Fermi energy: $\xi_{\vc{k}}=\frac{|\vc{k}|^{2}-k_{\mrm{F}}^{2}}{2m}$,
with $k_{\mrm{F}}$ being the Fermi momentum,
and $\hat{k}_{\perp}=(k_{1}+ik_{2})/{|\vc{k}|}$.
The mean-field Hamiltonian is diagonalized by quasiparticle operators $\gamma_{\vc{k}}$,
which are related to the electron operators $c_{\vc{k}}$ by
\begin{equation}
  c_{\vc{k}}=u_{\vc{k}}\gamma_{\vc{k}}+v_{-\vc{k}}^{\ast}\dg{\gamma}_{-\vc{k}}.
\end{equation}
Here $u_{\vc{k}}$ and $v_{\vc{k}}$ are given by
\begin{align}
  u_{\vc{k}}&=\sqrt{\frac{E_{\vc{k}}+\xi_{\vc{k}}}{2E_{\vc{k}}}}, \\
  v_{\vc{k}}&=\frac{\Delta \hat{k}_{\perp}^{\ast}}{\sqrt{2E_{\vc{k}}(E_{\vc{k}}+\xi_{\vc{k}})}}.
\end{align}
The set of data $\{E_{\vc{k}}, u_{\vc{k}}, v_{\vc{k}}\}_{\vc{k}}$ fully characterizes the chiral $p$-wave superfluid phase.

The spectral gap of particle-hole excitations closes at $\vc{k}=\pm 2k_{\mrm{F}}\vc{e}_{3}$ with $\vc{e}_{3}=(0,0,1)$.
In Appendix\ref{sec:app_chiralp}, the dynamic structure factor for small $\omega$ at this wavenumber is calculated as
\begin{equation}
  \frac{1}{V}S(2k_{\rm{F}}\vc{e}_{3},\omega)
	=\frac{mk_{\mrm{F}}\omega^{2}}{32\pi \Delta^{2}}.
\end{equation}
Thus the integral
\begin{equation}
  \frac{1}{N}\int_{0}^{\infty}\frac{S(2k_{\rm{F}}\vc{e}_{3},\omega)}{\omega^{3}}d\omega
\end{equation}
diverges to infinity at the thermodynamic limit.
Thus we again recognize the necessity to treat nonuniform potential nonperturbatively
to investigate superfluid property of chiral $p$-wave superfluid.

\section{Conclusion}
\label{sec:conclusion}
In this paper,
we constructed perturbation theory for superfluid under nonuniform potential
to discuss the superfluid properties of general systems.

In Sect. \ref{sec:definition}, we defined the superfluid fraction, which indicates
the extent to which the system shows NCRI.
By definition, the superfluid fraction gets trivially unity without nonuniform potential,
which of course does not mean that any system can be superfluid
but just reflects the fact that a rotating container is the same as a stationary one
in absence of nonuniform potential.
Thus we have to treat a system with nonuniform potential 
to discuss superfluidity.

In Sect. \ref{sec:perturbation}, we derived the perturbation expansion of the superfluid fraction
in terms of the strength of nonuniform potential.
We saw that the coefficient of the leading order reflects
the property of the system only through the dynamic structure factor.
We also derived the perturbation expansion of the supefluid fraction
under presence of random potential.

In Sect. \ref{sec:bounds}, we derived upper and lower bounds
for the superfluid fraction perturbatively by estimating the coefficient.
Both the upper and lower bounds contain the static structure factor,
and the lower bound also reflects the character of the system
through the excitation spectrum.

In Sect. \ref{sec:examples}, we investigated some examples.
For ideal bosons under periodic potential,
the coefficient converges for nonzero wavenumber implying superfluidity of the ground state is robust against nonuniform potential,
but diverges at the long wavelength limit.
For interacting bosons under periodic potential,
we saw the superfluidity is stable for arbitrary wavenumber
with help of the single mode approximation.
As a special case of interacting bosons, we treated Bose gas in the mean-field limit,
in which excited states admit the picture of Bogoliubov quasiparticles,
and saw its superfluidity is robust under random potential.
Moreover we found stronger particle interaction implies larger superfluid fraction at perturbative level.
We also confirmed for the superfluidity of ideal fermions and Fermi liquid,
which have large density fluctuation at low energy, the perturbation coefficients diverge.
The superfluidity of TLL is confirmed to be robust only when
the Luttinger paramater $K$ is larger than $2$.
We cannot make any conclusion about superfluid property of TLL with $1<K\le 2$ in our perturbative approach.
The final example was chiral $p$-wave superfluid, which has point nodes.
From the dependence of the dynamic structure factor on $\omega$,
we saw that the perturbation coefficient diverges on the spectral nodes.

We did not investigate some interesting examples.
As one of them, we mention supersolid~\cite{Leggett1970, WB2012, Saslow2012},
which is a phase with spontaneously broken translational symmetry,
and in which diagonal and off-diagonal long range orders coexist.
It is said to exhibit superfluidity due to its off-diagonal long range order,
but it is interesting to investigate the superfluid property of this system
based on our formulation.

In discussions by Leggett~\cite{Leggett1970, Leggett1998},
it is suggested that superfluid fraction is suppressed
if the particle density is nonuniform.
Thus it is expected that superfluidity is broken by nonuniform potential,
if the system has large density fluctuation,
which allows the ground state adjusting to the nonuniform potential
to be constructed as superposition of low energy states.
In this paper, we directly connected these two concepts, superfluidity and the density fluctuation.
Also we assumed nothing about BEC, thus our results are applicable to arbitrary systems.

Finally we make a comment on the relation between superfluidity and BEC.
Since BEC is neither necessary nor sufficient to superfluidity of the ground state,
it is natural that BEC seems have nothing to do with superfluidity in our formulation.
Nevertheless, we can expect BEC gives mechanism to suppress the density fluctuation,
because if the ground state is Bose-Einstein condensate,
we can expect that a process annihilating condensed one-particle state and creating quasiparticle state
contributes dominantly to the matrix elements in the dynamic structure factor
and that the single-mode approximated structure factor well describes
the density fluctuation.
How this picture works in realistic settings may be one of key issues
in the future study of superfluidity.

\begin{acknowledgments}
\acknowledgments

\section*{Acknowledgment}

The authors are grateful to M. Kunimi
and Y. Masaki for fruitful discussion
and encouragement.


\end{acknowledgments}

\begin{appendix}
\appendix
\section{Coefficients of $\kappa^{0}$ and $\kappa^{1}$}
\label{sec:appendixa}
In this appendix, we show the coefficients of $\kappa^{0}$ and $\kappa^{1}$
in the series in Eq. (\ref{eq:c2series}) actually vanish.

The coefficient of $\kappa^{0}$ is
\begin{equation}
  \frac{1}{2\pi i}\int_{K_{\epsilon}}r_{0}(E) dE=\frac{1}{2\pi i}\int_{K_{\epsilon}}\frac{g_{0}(E)}{E_{0}-E} dE,
\end{equation}
with $g_{0}$ being given by
\begin{equation}
  g_{0}(E)=\braket{Q_{0}\Omega_{0},P_{d}(H_{0}-E)^{-1}P_{d}Q_{0}\Omega_{0}}.
\end{equation}
Since
\begin{equation}
  Q_{0}=-\frac{1}{2\pi i}\int_{K_{\pr{\epsilon}}}(H_{0}-E)^{-1}dE
\end{equation}
is the projection onto the one dimensional subspace spanned by the ground state $\Omega_{0}$,
we have $Q_{0}\Omega_{0}=\Omega_{0}$.
Thus we obtain
\begin{equation}
  g_{0}(E)=\braket{\Omega_{0},P_{d}(H_{0}-E)^{-1}P_{d}\Omega_{0}}.
\end{equation}
Here $\Omega_{0}$ is the ground state of $H_{0}$, which commutes with the momentum $P_{d}$,
then $\Omega_{0}$ is also an eigenstate of the momentum.
Also we know that the momentum of the unique ground state is zero, thus $P_{d}\Omega_{0}=0$.
From this fact, the function $g_{0}(E)$ is identically zero,
and we can see the coefficient of $\kappa^{0}$ vanishes.

Next we calculate the coefficient of $\kappa^{1}$ as follows.
\begin{align}
  \frac{1}{2\pi i}\int_{K_{\epsilon}}r_{1}(E) dE
  	&=\frac{1}{2\pi i}\int_{K_{\epsilon}}\frac{1}{E_{0}-E}(g_{1}(E)-f_{1}r_{0}(E)) dE  \notag \\
  	&=\frac{1}{2\pi i}\int_{K_{\epsilon}}\frac{g_{1}(E)}{E_{0}-E}dE.
\end{align}
We used that $r_{0}(E)=0$ in the second equality.
$g_{1}(E)$ is 
\begin{equation}
  g_{1}(E)	
  =\sum_{i=0}^{1}\sum_{j=0}^{1-i}(-1)^{j}\braket{Q_{i}\Omega_{0},P_{d}T^{j}_{H_{0},U}P_{d}Q_{1-i-j}\Omega_{0}}.
\end{equation}
In this summation, either $Q_{i}$ or $O_{1-i-j}$ has to be $Q_{0}$, 
which leads to $P_{d}Q_{i}\Omega_{0}=P_{d}\Omega_{0}=0$ or $P_{d}Q_{1-i-j}\Omega_{0}=0$, respectively.
Then we can see $g_{1}(E)=0$ identically and the coefficient of $\kappa^{1}$ also vanishes.

\section{Structure factor of Tomonaga-Luttinger liquid}
\label{sec:app_tll}
In this appendix, we compute the dynamic structure factor of TLL
and especially investigate its low energy behavior at $k=2k_{\mrm{F}}$.

We start from the imaginary time density-density correlation function of TLL given by\cite{Giamarchi2004}
\begin{align}
  &\braket{\Omega_{0},T_{\tau}[\rho(x,\tau)\rho(0,0)]\Omega_{0}} \notag \\
  	&=\rho^{2}+\frac{K}{2\pi^{2}}\frac{(u\tau)^{2}-x^{2}}{(x^{2}+(u\tau)^{2})^{2}}
  		+\rho^{2}A_{2}\cos(2k_{\rm{F}}x)\left(\frac{\alpha}{r}\right)^{2K} \notag \\
  	&\hspace{30pt}+\rho^{2}A_{4}\cos(4k_{\rm{F}}x)\left(\frac{\alpha}{r}\right)^{8K}+\cdots.
\end{align}
Here $r=\sqrt{x^{2}+(c\tau)^{2}}$ and $\rho$ is the mean particle density $\rho=\braket{\Omega_{0},\rho(x,t)\Omega_{0}}$.
The part $\cdots$ above is a collection of contributions from other primary fields.

We obtain the real time correlation function $Q^{T}(x,t)=\braket{\Omega_{0},T[\rho(x,t)\rho(0,0)]\Omega_{0}}$ by setting
\begin{equation}
  \tau = it +\epsilon \sgn (t).
\end{equation}
The above correlation function is a time ordered one.
The correlation function without time ordering is obtained as
\begin{equation}
  S(x,t):=\braket{\Omega_{0},\rho(x,t)\rho(0,0)\Omega_{0}}=
  	\begin{cases}
  	  Q^{T}(x,t) & t \ge 0 \\
  	  Q^{T}(x,t)^{\ast} & t < 0.
  	\end{cases}
\end{equation}

The dynamic structure factor $S(k,\omega)$ is the Fourier transform of this density-density correlation:
\begin{equation}
  S(k,\omega)=\frac{L}{2\pi} \int S(x,t) e^{-i(kx-\omega t)}dxdt.
\end{equation}
Here $L$ denotes the length of the system.
The integral of $x$ is over a finite length,
but we extend it to an integral over whole real numbers
assuming it does not change essential properties.
The constant term in $S(x,t)$ leads to delta functions.
The Fourier transform of the second term is exactly calculated.
The second term is decomposed as
\begin{equation}
  S^{(1)}(x,t)
  	= -\frac{K}{(2\pi)^{2}}\left[ \frac{1}{(x+ct-i\epsilon)^{2}}+\frac{1}{(x-ct+i\epsilon)^{2}}\right]
\end{equation}
Let us write the Fourier transform as
\begin{equation}
  S^{(1)}(k,\omega)=\frac{L}{2\pi}\int S^{(1)}(x,t)e^{-i(kx-\omega t)}dxdt.
\end{equation}
We extend this integral to an integral in the complex plane,
then the integrand has poles of degree $2$ at $\pm (ut \mp i\epsilon)$.
We only focus on the case $k>0$ and add an integral contour in the lower half plane to obtain
\begin{align}
  S^{(1)}(k,\omega)=LKk\delta(\omega-ck).
\end{align}

The third term is
\begin{equation}
  S^{(2)}(x,t)=\frac{\rho^{2}A_{2}}{2}\frac{e^{i2k_{\rm{F}}x}+e^{-i2k_{\rm{F}}x}}{((x-ct+i\epsilon)(x+ct-i\epsilon))^{K}}.
\end{equation}

We write the Fourier transform of
\begin{equation}
  \frac{e^{\pm i2k_{\rm{F}}x}}{(x-ct+i\epsilon)^{K}(x+ct-i\epsilon)^{K}}
\end{equation}
as
\begin{equation}
  F_{\pm}(k,\omega):= \int dx dt \frac{e^{i(\pm 2k_{\rm{F}}-k)x+i\omega t}}{(x-ct+i\epsilon)^{K}(x+ct-i\epsilon)^{K}}.
\end{equation}
Then the third term contributes to the dynamic structure factor as the following form:
\begin{equation}
  S^{(2)}(k,\omega)=\frac{L\rho^{2}A_{2}}{4\pi}(F_{+}(k,\omega)+F_{-}(k,\omega)).
\end{equation}

Let us calculate $F_{\pm}(k,\omega)$.
To do it, we introduce the light-cone coordinates by
\begin{equation}
  \xi_{\pm}:=x\pm ct,
\end{equation}
and transform the integral variable from $x$ and $t$ to $\xi_{+}$ and $\xi_{-}$:
\begin{equation}
  F_{\pm}(k,\omega)=\frac{1}{2c}\int d\xi_{+}d\xi_{-}
  	\frac{e^{i(\pm k_{\rm{F}}-k/2+\omega/2c)\xi_{+}}e^{i(\pm k_{\rm{F}}-k/2-\omega/2c)\xi_{-}}}{(\xi_{+}-i\epsilon)^{K}(\xi_{-}+i\epsilon)^{K}}.
\end{equation}
We further introduce new integral variable $\zeta_{\pm}$ as follows:
\begin{align}
  \left(\pm k_{\rm{F}}-\frac{k}{2}+\frac{\omega}{2c}\right)\xi_{+}&=\sgn\left(\pm k_{\rm{F}}-\frac{k}{2}+\frac{\omega}{2c}\right)\zeta_{+}, \\
  \left(\pm k_{\rm{F}}-\frac{k}{2}-\frac{\omega}{2c}\right)\xi_{-}&=\sgn\left(\pm k_{\rm{F}}-\frac{k}{2}-\frac{\omega}{2c}\right)\zeta_{-}.
\end{align}
Then the integral becomes
\begin{align}
&F_{\pm}(k,\omega) \notag \\
	&=\frac{1}{2c}\Big|\pm k_{\rm{F}}-\frac{k}{2}+\frac{\omega}{2c}\Big|^{K-1}\Big|\pm k_{\rm{F}}-\frac{k}{2}-\frac{\omega}{2c}\Big|^{K-1} \notag \\
  	&\hspace{10pt}\times\left(\int d\zeta_{+}\frac{e^{i\sgn(\pm k_{\rm{F}}-k/2+\omega/2c)\zeta_{+}}}{(\zeta_{+}-i\epsilon)^{K}}\right)
  	\left(\int d\zeta_{-}\frac{e^{i\sgn(\pm k_{\rm{F}}-k/2-\omega/2c)\zeta_{-}}}{(\zeta_{-}-i\epsilon)^{K}}\right)
\end{align}

Since the integrand of the $\zeta_{+}$ integral is analytic in the lower half plane, it is proportional to $\theta(\pm k_{\rm{F}}-k/2+\omega/2c)$.
We write it as
\begin{equation}
  \int d\zeta_{+}\frac{e^{i\sgn(\pm k_{\rm{F}}-k/2+\omega/2c)\zeta_{+}}}{(\zeta_{+}-i\epsilon)^{K}}
  	=\theta\left(\pm k_{\rm{F}}-\frac{k}{2}+\frac{\omega}{2c}\right)c_{\pm}^{+}(K).
\end{equation}
In the same way the $\zeta_{-}$ integral can be written as
\begin{equation}
  \int d\zeta_{-}\frac{e^{i\sgn(\pm k_{\rm{F}}-k/2-\omega/2c)\zeta_{+}}}{(\zeta_{-}+i\epsilon)^{K}}
  	=\theta\left(\mp k_{\rm{F}}+\frac{k}{2}+\frac{\omega}{2c}\right)c_{\pm}^{-}(K).
\end{equation}

Thus we finally obtain
\begin{align}
  &S^{(2)}(k,\omega) \notag \\
  &=\frac{L\rho^{2}A_{2}}{8\pi c}\Biggl[ \Big|k_{\rm{F}}-\frac{k}{2}+\frac{\omega}{2c}\Big|^{K-1}\Big|k_{\rm{F}}-\frac{k}{2}-\frac{\omega}{2c}\Big|^{K-1}c_{+}^{+}(K)c_{+}^{-}(K) \notag \\
  &\hspace{50pt}	\times\theta\left(k_{\rm{F}}-\frac{k}{2}+\frac{\omega}{2c}\right)\theta\left(- k_{\rm{F}}+\frac{k}{2}+\frac{\omega}{2c}\right) \notag \\
  &\hspace{30pt}+\Big|-k_{\rm{F}}-\frac{k}{2}+\frac{\omega}{2c}\Big|^{K-1}\Big|-k_{\rm{F}}-\frac{k}{2}-\frac{\omega}{2c}\Big|^{K-1}c_{-}^{+}(K)c_{-}^{-}(K) \notag \\
  &\hspace{50pt}	\times\theta\left(-k_{\rm{F}}-\frac{k}{2}+\frac{\omega}{2c}\right)\theta\left(k_{\rm{F}}+\frac{k}{2}+\frac{\omega}{2c}\right) \Biggr].
\end{align}

The low energy contribution of the dynamic structure factor at $k=2k_{\mrm{F}}$
comes from $S^{(2)}$, thus we have
\begin{equation}
	S(k=2k_{\mrm{F}},\omega)\propto L\omega^{2(K-1)}
\end{equation}
at small $\omega$.

\section{Structure factor of spinless chiral $p$-wave superfluid}
\label{sec:app_chiralp}
In this appendix, we investigate the low energy behavior of
the dynamic structure factor of chiral $p$-wave superfluid
at $\vc{k}=2k_{\mrm{F}}\vc{e}_{3}$.

Let us first calculate the matrix element $\braket{\Omega_{n},\rho_{-\vc{k}}\Omega_{0}}$,
with $\rho_{-\vc{k}}=\sum_{\vc{q}}\dg{c}_{\vc{q}+\vc{k}}c_{\vc{q}}$,
and $c_{\vc{k}}^{(\dagger)}$ being electron operators.
For $\vc{k}\neq \vc{0}$, it is
\begin{equation}
  \braket{\Omega_{n},\rho_{-\vc{k}}\Omega_{0}}
	=\sum_{\vc{q}}u_{\vc{q}+\vc{k}}^{\ast}v_{-\vc{q}}^{\ast}\braket{\Omega_{n},\dg{\gamma}_{\vc{q}+\vc{k}}\dg{\gamma}_{-\vc{q}}\Omega_{0}}.
\end{equation}
Here we note the ground state $\Omega_{0}$ is characterized by $\gamma_{\vc{k}}\Omega_{0}=0$ for all $\vc{k}$.

Then the dynamic structure factor $S(\vc{q},\omega)$ is calculated as
\begin{align}
  &S(\vc{k},\omega) \notag \\
  	&=\sum_{n}|\braket{\Omega_{n},\rho_{-\vc{k}}\Omega_{0}}|^{2}\delta(\omega-\omega_{n0}) \notag \\
  	&=\sum_{\vc{q}}\sum_{n}\braket{\Omega_{0},\rho_{\vc{k}}\Omega_{n}}u_{\vc{q}+\vc{k}}^{\ast}v_{-\vc{q}}^{\ast}\braket{\Omega_{n},\dg{\gamma}_{\vc{q}+\vc{k}}\dg{\gamma}_{-\vc{q}}\Omega_{0}}\delta(\omega-\omega_{n0}).
\end{align}
For a fixed $\vc{q}$, the only nonvanishing term in the $n$-summation comes from the excited state $\Omega_{n}=\dg{\gamma}_{\vc{q}+\vc{k}}\dg{\gamma}_{-\vc{q}}\Omega_{0}$.
Thus the calculation goes as
\begin{align}
  &S(\vc{k},\omega) \notag \\
  	&=\sum_{\vc{q}}u_{\vc{q}+\vc{k}}^{\ast}v_{-\vc{q}}^{\ast}\braket{\Omega_{0},\rho_{\vc{k}}\dg{\gamma}_{\vc{q}+\vc{k}}\dg{\gamma}_{-\vc{q}}\Omega_{0}}
  		\delta(\omega-(E_{\vc{q}+\vc{k}}+E_{-\vc{q}})) \notag \\
	&=\sum_{\vc{q}}(|u_{\vc{q}}v_{\vc{k}-\vc{q}}|^{2}-u_{\vc{k}-\vc{q}}v_{\vc{q}}u_{\vc{q}}^{\ast}v_{\vc{k}-\vc{q}}^{\ast})\delta(\omega-(E_{\vc{q}}+E_{\vc{k}-\vc{q}})).
\end{align}
At the thermodynamic limit, we have
\begin{align}
  &\frac{1}{V}S(\vc{k},\omega) \notag \\
  &\to \int \frac{d\vc{q}}{(2\pi)^{3}}
	(|u_{\vc{q}}v_{\vc{k}-\vc{q}}|^{2}-u_{\vc{k}-\vc{q}}v_{\vc{q}}u_{\vc{q}}^{\ast}v_{\vc{k}-\vc{q}}^{\ast})\delta(\omega-(E_{\vc{q}}+E_{\vc{k}-\vc{q}})).
\end{align}

Next we calculate this integral and investigate the $\omega$ dependence of the dynamic structure factor at $\vc{k}=2k_{\rm{F}}\vc{e}_{3}$,
where the spectral gap of particle-hole excitation closes.
In the following, we fix $\vc{k}=2k_{\rm{F}}\vc{e}_{3}$.
Then $\vc{q}$ satisfying $\omega-(E_{\vc{q}}+E_{\vc{k}-\vc{q}})=0$ is around $(0,0,k_{\rm{F}})$.
Thus we introduce a new variable $\tilde{\vc{q}}$ such that
\begin{equation}
  \vc{q}=(0,0,k_{\rm{F}})+\tilde{\vc{q}}.
\end{equation}
Then $\vc{k}-\vc{q}=(0,0,k_{\rm{F}})-\tilde{\vc{q}}$.
This allows one to find, regarding $E_{\vc{q}}$ as a function of $\tilde{\vc{q}}$:
\begin{equation}
  E_{\vc{q}}=\epsilon(\tilde{\vc{q}}),
\end{equation}
we can write $E_{\vc{k}-\vc{q}}=\epsilon(-\tilde{\vc{q}})$.

To calculate the integral, we make two approximations.
In the one particle excitation spectrum, $\hat{q}_{\perp}$ is normalized as
\begin{equation}
  \hat{q}_{\perp}=(q_{1}+iq_{2})/|\vc{q}|.
\end{equation}
As the first approximation of the coherence factor,
we replace the normalizing $|\vc{q}|$ by $k_{\mrm{F}}$.
Secondly, we linearlize $\xi_{\vc{q}}$ around $(0,0,k_{\rm{F}})$ as
\begin{equation}
  \xi_{\vc{q}}=\frac{|\vc{q}|^{2}}{2m}-\frac{k_{\rm{F}}^{2}}{2m}\simeq \frac{k_{\rm{F}}}{m}\tilde{q}_{3}.
\end{equation}
Here $\tilde{q}_{3}$ is the third component of $\tilde{\vc{q}}$ introduced in above.
Thus we obtain one-particle excitation spectrum around $(0,0,k_{\rm{F}})$ as
\begin{equation}
  E_{\vc{q}}=\epsilon(\tilde{\vc{q}})=\sqrt{\left(\frac{k_{\rm{F}}}{m}\tilde{q}_{3}\right)^{2}+\left(\frac{\Delta}{k_{F}}\right)^{2}|\tilde{q}_{\perp}|^{2}}.
\end{equation}
It is soon be recognized that $E_{\vc{k}-\vc{q}}=\epsilon(-\tilde{\vc{q}})=\epsilon(\tilde{\vc{q}})$.
Setting newly $\eta_{1}=\tilde{q}_{1}/k_{\mrm{F}}, \eta_{2}=\tilde{q}_{2}/k_{\mrm{F}}$ and $\eta_{3}=\frac{k_{\rm{F}}}{m\Delta}\tilde{q}_{3}$,
we can simply write
\begin{equation}
  \epsilon(\tilde{\vc{q}})=\Delta\sqrt{\eta_{1}^{2}+\eta_{2}^{2}+\eta_{3}^{2}}=:\Delta\eta.
\end{equation}
We use the same approximation to $u_{\vc{q}}$ and $v_{\vc{q}}$.
Then we get
\begin{equation}
  u_{\vc{q}}=\sqrt{\frac{\eta+\eta_{3}}{2\eta}},\hspace{20pt} v_{\vc{q}}=\frac{\eta_{\perp}^{\ast}}{\sqrt{2\eta(\eta+\eta_{3})}},
\end{equation}
and
\begin{equation}
  u_{\vc{k}-\vc{q}}=\sqrt{\frac{\eta-\eta_{3}}{2\eta}},\hspace{20pt} v_{\vc{k}-\vc{q}}=\frac{-\eta_{\perp}^{\ast}}{\sqrt{2\eta(\eta-\eta_{3})}}.
\end{equation}

Following these preliminaries, the coherence factor becomes
\begin{equation}
  |u_{\vc{q}}v_{\vc{k}-\vc{q}}|^{2}-u_{\vc{k}-\vc{q}}v_{\vc{q}}u_{\vc{q}}^{\ast}v_{\vc{k}-\vc{q}}^{\ast}
 =\frac{\eta_{\perp}^{2}}{2\eta(\eta-\eta_{3})}.
\end{equation}
Then the dynamic structure factor at $k=2k_{\mrm{F}}\vc{e}_{3}$ becomes
\begin{align}
	\frac{1}{V}S(2k_{\mrm{F}}\vc{e}_{3},\omega)
		&=\frac{mk_{\mrm{F}}\Delta}{(2\pi)^{3}}\int d\vc{\eta}\frac{\eta_{\perp}^{2}}{2\eta(\eta-\eta_{3})}\delta(\omega-2\Delta\eta) \notag \\
		&=\frac{mk_{\mrm{F}}\omega^{2}}{32\pi\Delta^{2}}.
\end{align}
\end{appendix}





\bibliographystyle{jpsj}
\bibliography{reference} 


\end{document}